\documentclass[journal]{IEEEtran}
\usepackage[switch]{lineno}

\usepackage{tikz}
\usepackage{textcomp}
\usepackage{hyperref}

\usepackage{color}
\usepackage{amsmath,amsfonts}
\usepackage{algorithmic}
\usepackage{algorithm}
\usepackage{array}
\usepackage[caption=false,font=normalsize,labelfont=sf,textfont=sf]{subfig}
\usepackage{textcomp}
\usepackage{stfloats}
\usepackage{url}
\usepackage{verbatim}
\usepackage{graphicx}
\usepackage{cite}
\definecolor{green}{RGB}{0,180,0}
\definecolor{brown}{RGB}{160,82,45}
\ifCLASSINFOpdf
\else
\fi
\newcommand\copyrighttext{%
  \footnotesize \textcopyright~2024 IEEE. Personal use of this material is permitted. Permission from IEEE must be obtained for all other uses, in any current or future media, including reprinting/republishing this material for advertising or promotional purposes, creating new collective works, for resale or redistribution to servers or lists, or reuse of any copyrighted component of this work in other works.
  DOI: \href{https://ieeexplore.ieee.org/document/10316302}{10.1109/TIM.2023.3332388}}
\newcommand\copyrightnotice{%
\begin{tikzpicture}[remember picture,overlay]
\node[anchor=south,yshift=2pt] at (current page.south) {\fbox{\parbox{\dimexpr\textwidth-\fboxsep-\fboxrule\relax}{\copyrighttext}}};
\end{tikzpicture}%
}

\begin{document}


%
\title{Performance Analysis of Sequential Carrier- and Code-Tracking Receivers in the Context of High-Precision Space-Borne Metrology Systems}
%
%
%

\author{Philipp~Euringer,
        Gerald~Hechenblaikner,
        Francis~Soualle,
        and~Walter~Fichter
\thanks{This work was supported by funding from the Max-Planck-Institut für Gravitationsphysik (Albert-Einstein-Institut), based on a grant by the Deutsches Zentrum für Luft- und Raumfahrt (DLR) and by the Bundesministerium für Wirtschaft und Klimaschutz based on a resolution of the German Bundestag (Project Ref. Number 50 OQ 1801) \textit{(Corresponding author: Philipp Euringer.)}}
\thanks{Philipp Euringer and Gerald Hechenblaikner are with Airbus Space Systems, Airbus Defence and Space GmbH, Claude-Dornier-Straße, 88090 Immenstaad am Bodensee, Germany (e-mail: philipp.euringer@airbus.com).}
\thanks{Francis Soualle is with Airbus Space Systems, Airbus Defence and Space GmbH, Willy-Messerschmitt-Straße 1, 82024 Taufkirchen, Germany.}
\thanks{Walter Fichter is with the Institute of Flight Mechanics and Control, University of Stuttgart, Pfaffenwaldring 27, 70569 Stuttgart, Germany.}}

%
%

\markboth{EURINGER \MakeLowercase{\textit{et al.}}: PERFORMANCE ANALYSIS OF SEQUENTIAL CARRIER- AND CODE-TRACKING RECEIVERS}%
{Performance analysis of sequential carrier- and code-tracking receivers in the context of high-precision space-borne metrology systems}
%




\maketitle
\copyrightnotice

\begin{abstract}
Future space observatories achieve detection of gravitational waves
by interferometric measurements of a carrier phase, allowing to determine
relative distance changes, in combination with an absolute distance
measurement based on the transmission of pseudo-random noise chip
sequences. In addition, usage of direct-sequence spread spectrum modulation
enables data transmission. Hereafter, we report on the findings of
a novel performance evaluation of planned receiver architectures,
performing phase and distance readout sequentially, addressing the
interplay between both measurements. An analytical model is presented
identifying the power spectral density of the chip modulation at frequencies
within the measurement bandwidth as the main driver for phase noise.
This model, verified by numerical simulations, excludes binary phase-shift
keying modulations for missions requiring pico-meter noise levels
at the phase readout, while binary offset carrier modulation, where
most of the power has been shifted outside the measurement bandwidth,
exhibits superior phase measurement performance. Ranging analyses
of the delay-locked loop reveal strong distortion of the pulse shape
due to the preceding phase tracking introducing ranging bias variations.
Numerical simulations show that these variations, however, which originate
from data transitions, are compensated by the delay tracking loop,
enabling sub-meter ranging accuracy, irrespective of the modulation
type.\\
\end{abstract}


%

\section{Introduction}

\subsection{High-precision space-borne metrology systems\label{subsec:LISA-specific}}

Recent years have seen rapid advancements in missions using optical
interferometry in space. Planned missions, such as the space observatories
LISA \cite{danzmann2003lisa} (Europe/US), TAIJI \cite{luo2020taiji}
and TIANQIN \cite{mei2021tianqin} (China), and DECIGO \cite{kawamura2021current}
(Japan), aim to detect gravitational waves by interferometric measurements
across huge distances of up to several million kilometers in order
to achieve the required strain sensitivity on the order of 1 part
in $10^{20}$.

LISA is a planned ESA/NASA mission, currently in Phase $B1$ of the
development, with a constellation of 3 spacecraft (SC) forming an
equilateral triangle of 2.5 million km arm-length. The measurement
methodology is as follows. Two one-way optical links are established
in opposing directions between each pair of SC. These are primarily
used for heterodyne interferometric measurements of the carrier phase
to determine relative distance changes with an accuracy of approximately
10 pm$/\sqrt{\text{Hz}}$ in a measurement bandwidth from 0.1 mHz
to 1 Hz \cite{danzmann2011LISA}. In addition, accurate knowledge
of the inter-SC distance is needed for a post-processing technique
referred to as time-delay-interferometry \cite{tinto2021time}. Thereby,
virtual Michelson interferometers are synthesized from individual
arm measurements in order to suppress laser frequency noise \cite{muratore2020revisitation}
and tilt-to-length coupling noise \cite{houba2022lisa} by several
orders of magnitude. Therefore, as a secondary function, the links
also allow determining the absolute distance (ranging) and exchanging
data in between SC by modulating pseudo-random noise (PRN) sequences
onto the carrier and data bits onto the PRN code sequences \cite{heinzel2011auxiliary,sutton2010laser}.
Knowledge of the absolute distance is obtained by correlating the
received PRN code sequence with a local SC replica, which yields the
relative code delays and the associated inter-SC distance within an
ambiguity range, in a similar way as done in the radio frequency
domain by global navigation satellite systems (GNSS). The ambiguity
can then be resolved by a combination of radio-frequency ranging measurements
from ground stations and orbit prediction between measurements \cite{danzmann2011LISA}.
One of the promising detection architectures, facilitating this measurement
methodology, employs a phase-locked loop (PLL), used for measuring
the interferometric phase, sequentially followed by a delay-lock-loop
(DLL), used for measuring the code delay to determine the range \cite{Esteban:11,Delgado2012}.

\subsection{Related work}

In order to validate the measurement methodology including the detection
architecture\textcolor{brown}{{} }great effort has been made to identify
noise sources and evaluate the phase measurement and ranging accuracy.
Previous studies mainly focused on phase noise analysis, with significant
experimental work conducted in this domain \cite{Shaddock2006, Liang2021, Li2020,Liu2014, Gerberding2015}.
Based on theoretical and experimental evaluation, dominating noise
contributions in the detection chain have been identified, in particular
shot noise \cite{Meers1991,Niebauer1991}, laser intensity noise \cite{Hechenblaikner:13,Wissel2022,Wissel2023},
and sampling jitter and thermal drift \cite{Liang2015}, while alternative
designs have been proposed minimizing these effects \cite{Ales2015, Liang2018}.
In addition, performance evaluations on analytical and experimental
basis considering the ranging accuracy have been conducted \cite{Esteban2009, sutton2010laser, Sutton_2013, heinzel2011auxiliary, Xie2023}.
It should be noted that the authors of \cite{sutton2010laser} have
been the first to discuss the impact of the PLL on the DLL ranging
performance and to highlight the importance of the modulation scheme.
\\
However, only a small number of publications consider both the phase
measurement and the ranging accuracy \cite{Esteban:11,Delgado2012},
and these publications are mainly limited to experimental findings,
while a parametric evaluation of both measurement principles is absent.
In fact, a holistic evaluation of phase measurement and ranging accuracy
is indispensable as both parameters affect the feasibility of the
measurement methodology. Moreover, both measurements may affect each
other, via signal modulation and processing and thus phase measurement
performance may be antagonistic to ranging accuracy.

\subsection{Major contributions}

In this article, we propose a theoretical foundation, unveiling the
interplay between phase measurement and ranging of sequential carrier-
and code-tracking receivers in the context of high-precision space-borne
metrology systems. A generic model is introduced specifically for
each core function, namely phase measurement and ranging, which, by
neglecting external noise sources, provides in-depth insight into
the performance resulting from the interplay of both measurements
during signal tracking.

The main contributions of this article are given as follows.
\begin{enumerate}
\item A novel measurement performance evaluation of sequential carrier-
and code-tracking receivers considered for high-precision space-borne
metrology systems is conducted on a theoretical level. Although, this
analysis evaluates the performance losses resulting solely from architectural
design choices the model is verified by using representative signals
for high-precision space-borne metrology systems, including the respective
noise sources, to compare analytical predictions to simulations and
previous experimental results.
\item The analytical model for phase noise reveals the compelling influence
of the power spectral density of the pulse modulation at frequencies
within the measurement bandwidth on the phase noise. This highlights
the importance of the pulse modulation type as a design parameter.
\item A semi-analytical model for the ranging accuracy reveals the impact
of the phase measurement performed as a first step on the accuracy
of the ranging performed as a second step. The model identifies data
modulation of PRN sequences in combination with the processing of
the phase measurement performed as a first step as driver for degradation
of the ranging performance.
\item Numerical validation of the models exposes that the simplest of the
previously considered modulation schemes (binary phase-shift keying,
\cite{Esteban:11, heinzel2011auxiliary}) degrades the primary phase
measurement accuracy to unacceptable levels for missions requiring
pico-meter noise levels at the phase readout, while performance is
fully recovered when adopting an alternative modulation scheme, namely
binary offset carrier modulation. Similar modulation schemes have
been proposed in the analysis focusing on the ranging performance
\cite{sutton2010laser, Delgado2012}, while in this paper, we also
assess the impact of the code modulation on the phase measurement
performance and introduce a comprehensive performance analysis combining
phase-readout and ranging.\textcolor{green}{{} }It is only through
a combined analysis, as given in this paper, that the mutual dependencies
and their parametric relationships become apparent.
\end{enumerate}
The structure of the paper will be as follows. In the subsequent
section, a generic signal and receiver architecture following the
measurement principle as delineated in subsection \ref{subsec:LISA-specific}
will be introduced. The focus of section \ref{sec:Carrier-tracking-and}
is on interferometric measurement performance. Thereby, the phase
noise resulting from the architecture and the signal composition will
be identified and modeled, yielding an analytic expression. This expression
will be applied to two typical modulation schemes and compared to
the result of a numerical breadboard setup in MATLAB. The ranging
accuracy is modeled and assessed in section \ref{sec:Code-tracking-and}.
Similarly to the phase measurement, an evaluation is performed for
the identical modulation schemes, where we find that both schemes
can achieve sub-meter ranging accuracies.

\section{Signal composition and receiver architecture\label{subsec:Theory-for-Basic}}

The signal model of (\ref{eq:input_signal}) will be used for the
performance assessment in the following sections. It has modulation
features that support an absolute distance measurement combined
with a high-precision but ambiguous carrier measurement and may represent
the output of a heterodyne detection pre-processing step, as detailed
in \cite{Esteban2009, EESA}.
\begin{equation}
s(t)=\cos\left(\omega t+m_{\text{prn}}\sum_{j=-\infty}^{\infty}d_{j}\sum_{i=0}^{N-1}c_{i}p(t-iT_{c}-jNT_{c})\right).\label{eq:input_signal}
\end{equation}
The first argument in the cosine represents the carrier phase of the
incoming\textcolor{brown}{{} }science signal, given by the angular frequency
$\omega$ and the time $t$. The second argument carries a chip sequence,
commonly known as PRN, enabling absolute ranging, where the chip sequence
consists of $N$ chips with a chip period $T_{c}$. Here, $c_{i}\in\left\{ \pm1\right\} $
and $p(t)$ represent the chip value of the $i$-th chip and the pulse
modulation, respectively. Thus, the pulse modulation can carry any
function in the range from $t$ to $t+T_{c}$, outside this range
it reads zero. In addition, the chip sequence is modulated with a
binary symbol ($d_{j}\in\left\{ \pm1\right\} $), with period $T_{s}=NT_{c}$,
for data transmission. Importantly, the modulation depth is controlled
by the parameter $m_{\text{prn}}$, known as modulation index. In
this sense, only parts of the carrier signal are modulated by the
chip sequence \cite{Delgado2012}. 

\begin{figure}[!t]
\centering{}\includegraphics[width=1\linewidth]{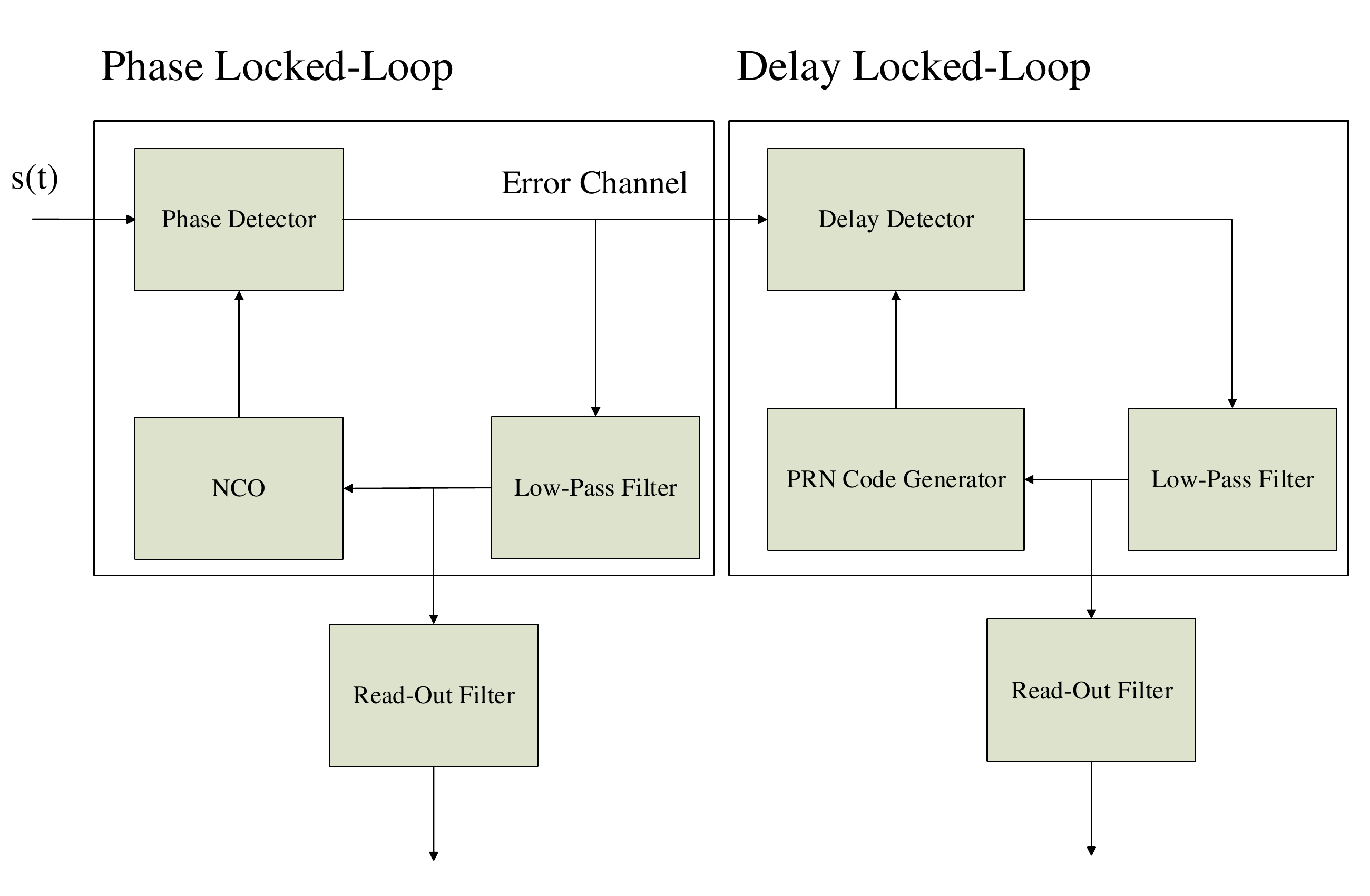}\caption{The incoming signal $s(t)$ is processed by an all-digital second-order
PLL, followed by a first-order DLL with a non-coherent early-late
discriminator for the ranging, and a prompt channel for data retrieval
(not shown). A similar model derived from control loop theory is applied
to both tracking loops with a detector, a low pass filter and a reference
signal generator. The output at the PLL read-out filter is a frequency
estimate, while the readout of the DLL is a chip rate estimate.
\label{fig:Generic-Receiver-Design}}
\end{figure}

Contrary to typical GNSS architectures, high-precision optical metrology
systems have a strong emphasis on the phase measurement, which requires
decoupling of code and phase estimation in order to avoid disturbance
effects of the DLL onto the PLL. This suggests using a sequential
PLL--DLL architecture, as proposed by Delgado \cite{Esteban:11, Delgado2012, EESA}.
The advantage of this cascaded architecture is a reduction of the
receiver complexity with separation of individual tracking functions
into single components. A generic model following this approach is
illustrated in Fig. \ref{fig:Generic-Receiver-Design}.
The receiver consists of two main components: the PLL and the DLL
being responsible for the carrier tracking, i.e. the phase readout,
and code tracking, i.e. the absolute ranging, respectively. The PLL
represents a second-order all-digital PLL consisting of a phase detector,
a first-order filter and a numerically controlled oscillator (NCO).
This generic design is studied extensively in several textbooks \cite{Stephens2002, Gardner2005},
and based on linear transfer models, it exhibits a low-pass filter
behavior in its closed-loop response, while the error response yields
a high-pass filter behavior \cite{Gardner2005}. Thus, the general
principle is as follows. Presuming that the error response bandwidth
of the PLL is smaller than the chip rate, the PLL is not able to track
chip fluctuations, and the chip sequences will remain in the error
channel of the PLL. Consequently, the error channel of the PLL serves
as an input to the DLL, cf. Fig. \ref{fig:Generic-Receiver-Design}.
The DLL estimates the time of arrival (TOA) of the incoming PRN sequence
based on an early-late discriminator, while a prompt channel is used
for data retrieval. The output of the early-late discriminator is
thereby low-pass filtered, yielding an update of the TOA for the PRN
code generator.

Finally, Table \ref{tab:Baseline-Parameter} lists parameters derived
from the generic model delineated in Fig. \ref{fig:Generic-Receiver-Design}.
Parameter values have been taken from tables listed in \cite{EESA}
and derived from \cite{Delgado2012}, and where necessary, values
not included were added to the table. Thus, these values are regarded
as relevant within high-precision space-borne metrology systems, in
particular LISA.
\begin{table}
\centering{}\caption{Baseline parameters\label{tab:Baseline-Parameter}}
\begin{tabular}{|l|l|l|}
\hline 
Parameter (Unit) & Symbol & Value\tabularnewline
\hline 
\hline 
Modulation index (-) & $m_{\text{prn}}$ & 0.1\tabularnewline
\hline 
Chip period (ms) & $T_{c}$ & 0.001\tabularnewline
\hline 
Sampling rate (MHz) & $f_{s}$ & 80\tabularnewline
\hline 
Symbol period (ms) & $T_{s}$ & 0.064\tabularnewline
\hline 
Carrier frequency (Mrad/s) & $\omega$ & 30$\pi$\tabularnewline
\hline 
PLL bandwidth (closed loop) (kHz) & $B_{\text{PLL}}$ & 250\tabularnewline
\hline 
DLL bandwidth (closed loop) (Hz) & $B_{\text{DLL}}$ & 10\tabularnewline
\hline 
Read-out filter PLL (Hz) & $B_{\text{F}}^{\text{p}}$ & 4\tabularnewline
\hline 
Read-out filter DLL (Hz) & $B_{\text{F}}^{\text{d}}$ & 10\tabularnewline
\hline 
BPSK early-late spacing ($T_{c}$) & $\Delta_{\text{BPSK}}$ & 0.5\tabularnewline
\hline 
BOC(1,1) early-late spacing ($T_{c}$) & $\Delta_{\text{BOC}}$ & 0.2\tabularnewline
\hline 
Wavelength at heterodyne detection (nm) & $\lambda$ & 1064\tabularnewline
\hline 
Responsivity (A/W) & $R_{\text{PD}}$ & 0.7\tabularnewline
\hline 
\end{tabular}
\end{table}

\section{Carrier tracking and phase read-out performance\label{sec:Carrier-tracking-and}}

\subsection{Signal modeling}

The measured carrier phase must be of the highest possible accuracy
in order to determine relative inter-SC distance changes. As shown
in (\ref{eq:input_signal}), the chip sequence will contribute to
the phase measurement noise. Thereby, the data symbol values are
not predefined and can be modeled as a random stream. In contrast,
the chip values are predefined and follow a fixed pattern. Nevertheless,
for sufficiently long PRN sequences, the chip stream will be considered
random for the sake of the following development. Consequently, both
variables may be modeled as Bernoulli variables $\hat{C},\hat{D}\in\{\pm1\}$,
motivating the introduction of a new Bernoulli variable $\hat{C}_{i}\hat{D}_{j}\rightarrow\hat{X}_{ij}\in\{\pm1\}$.
Thus, the expression for the resulting stochastic noise term $\hat{n}(t)$
is given by:
\begin{equation}
\hat{n}(t)=m_{\text{prn}}\sum_{n=-\infty}^{\infty}\hat{X}_{n}p(t-nT_{c}).\label{eq:noise_term_stochastic}
\end{equation}
Noting that the pulse modulation $p(t)$ is independent of the index
$n$, the former can be expressed according to:
\begin{align}
\hat{n}(t) & =m_{\text{prn}}\sum_{n=-\infty}^{\infty}\hat{X}_{n}\delta(t-nT_{c})\ast p(t),\label{eq:noise_term_stochastic_conv_extend}\\
 & =\hat{g}(t)\ast p(t),\label{eq:noise_term_stochastic_conv}
\end{align}
 where $\ast$ denotes the convolution operator. In the following
section, this term will be used for the calculation of the phase noise
measurement performance.

\subsection{Phase noise}

Phase noise is commonly measured as a power spectral density (PSD)
$S(f)$, where the variance $\sigma^{2}$ of phase noise can be deduced
from. Modeling the PLL as a linear time-invariant (LTI) system and
taking into account the processing according to Fig. \ref{fig:Generic-Receiver-Design},
the PSD at the readout is given by the noise power spectral density
$N_{\text{i}}(f)$ at the input of the PLL filtered by the closed-loop
transfer function of the PLL $H_{\text{PLL}}(f)$ and the read-out
filter $H_{\text{F}}^{\text{p}}(f)$ \cite{Gardner2005}. Since the
sampling rate (80 MHz) of the PLL is significantly larger than its
closed-loop bandwidth (250 kHz), the system can be well approximated
by a continuous representation in the frequency domain \cite{Stephens2002}.
Both, the closed-loop PLL and the read-out filter exhibit a low-pass
filter behavior and may be idealized according to $|H_{x}(f)|=\Pi_{-B_{x},B_{x}}(f)$.
Hereby, the boxcar function $\Pi(f)$ is defined via the Heaviside
step function $\theta$, according to: $\Pi_{a,b}(f)=\theta(f-a)-\theta(f-b)$.
Noting that the closed-loop bandwidth of the $B_{\text{PLL}}$ (250
kHz) exceeds the bandwidth of the read-out filter $B_{\text{F}}^{\text{p}}$
(Hz -- kHz) by orders of magnitude, results in a PSD and the corresponding
variance at the readout of:
\begin{align}
S(f) & =\Pi_{-B_{\text{PLL}},B_{\text{PLL}}}(f)\Pi_{-B_{\text{F}}^{\text{p}},B_{\text{F}}^{\text{p}}}(f)N_{\text{i}}(f)\approx\Pi_{-B_{\text{F}}^{\text{p}},B_{\text{F}}^{\text{p}}}(f)N_{\text{i}}(f),\label{eq:phase_noise_at_reatout}\\
\sigma^{2} & \approx\int_{-B_{\text{F}}^{\text{p}}}^{B_{\text{F}}^{\text{p}}}N_{\text{i}}(f)\text{d}f.\label{eq:variance_at_readout}
\end{align}
Thereby, (\ref{eq:phase_noise_at_reatout}) denotes the double-sided
PSD, which will be used in the remainder of this paper. In addition,
introducing the Fourier pairs $\hat{g}(t)\overset{\mathcal{F}}{\rightarrow}\hat{G}(f)$,
$p(t)\overset{\mathcal{F}}{\rightarrow}P(f)$ and $\hat{n}(t)\overset{\mathcal{F}}{\rightarrow}\hat{N}(f)$
and exploiting the convolution theorem on (\ref{eq:noise_term_stochastic_conv})
yields $\hat{N}(f)=\hat{G}(f)\cdot P(f)$. Thus the noise power spectral
density at the input of the PLL is given by:
\begin{align}
N_{\text{i}}(f) & =\underset{K\rightarrow\infty}{\text{lim}}\dfrac{1}{2KT_{c}}\langle|\hat{N}(f)|^{2}\rangle,\nonumber \\
 & =\dfrac{|P(f)|^{2}}{2T_{c}}\underset{K\rightarrow\infty}{\text{lim}}\dfrac{\langle|\hat{G}(f)|^{2}\rangle}{K},\nonumber \\
 & =\dfrac{|P(f)|^{2}}{2T_{c}}\underset{K\rightarrow\infty}{\text{lim}}\dfrac{m_{\text{prn}}^{2}}{K}\sum_{m,n=-K}^{K-1}\langle\hat{X}_{n}\hat{X}_{m}\rangle e^{\text{i}2\pi f(n-m)T_{c}},\nonumber \\
 & =m_{\text{prn}}^{2}\dfrac{|P(f)|^{2}}{T_{c}}\nonumber \\
 & \,\,\,\,\,\,\left[1+\underset{K\rightarrow\infty}{\text{lim}}\dfrac{1}{2K}\sum_{\underset{m\neq n}{m,n=-K}}^{K-1}\langle\hat{X}_{n}\hat{X}_{m}\rangle e^{\text{i}2\pi f(n-m)T_{c}}\right].\label{eq:noise_power_spectral_density_explicit}
\end{align}
Thereby, the brackets $\langle\rangle$, indicate the ensemble average
over the Bernoulli variables $\hat{X}_{n}$ and $\hat{X}_{m}$. If
$\hat{X}_{n}$ and $\hat{X}_{m}$ are uncorrelated for $n\neq m$,
the noise power spectral density is solely given by the PSD of the
pulse modulation multiplied by the modulation index squared. In this
limit, inserting (\ref{eq:noise_power_spectral_density_explicit})
into (\ref{eq:phase_noise_at_reatout}) and (\ref{eq:variance_at_readout})
yields:
\begin{align}
S(f) & =\Pi_{-B_{\text{F}}^{\text{p}},B_{\text{F}}^{\text{p}}}(f)m_{\text{prn}}^{2}\dfrac{|P(f)|^{2}}{T_{c}},\label{eq:phase_noise_density_approx}\\
\sigma^{2} & =m_{\text{prn}}^{2}\int_{-B_{\text{F}}^{\text{p}}}^{B_{\text{F}}^{\text{p}}}\dfrac{|P(f)|^{2}}{T_{c}}\text{d}f.\label{eq:variance_approx}
\end{align}
Equations (\ref{eq:phase_noise_density_approx}) and (\ref{eq:variance_approx})
highlight the impact of the PSD of the pulse modulation on the phase
noise performance, which is further quantified in the following section.

\subsection{Application and numerical verification\label{subsec:Application-and-numerical}}

\begin{figure}[!t]
\centering{}\includegraphics[width=1\linewidth]{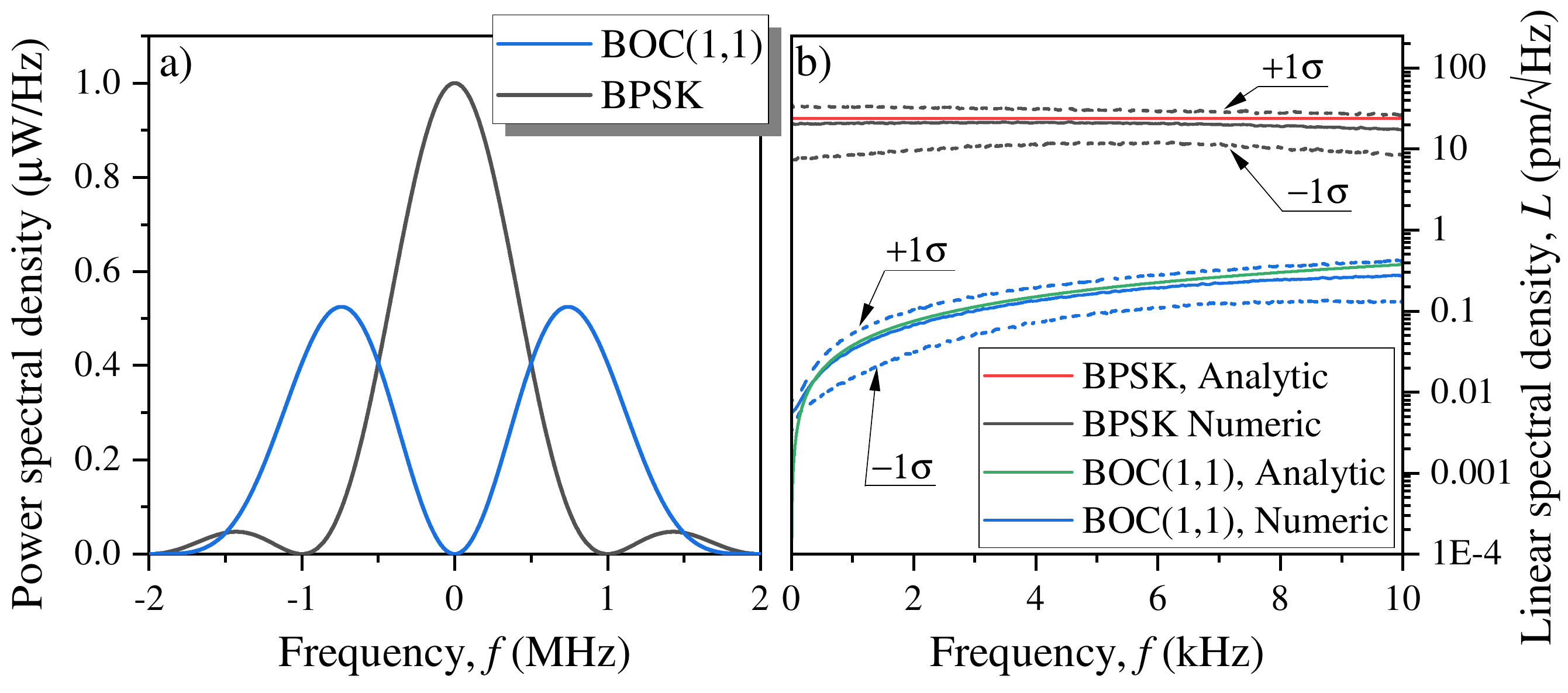}\caption{Panel a) illustrates the double-sided PSD for a BPSK (gray graph)
and BOC(1,1) (blue graph) modulated rectangular chip, exhibiting a
chip width of 0.001 ms. In panel b) the single-sided displacement
noise LSD of a numerical PLL simulation (gray lines for BPSK and blue
lines for BOC) and an analytical model (red line for BPSK and green
line for BOC) are compared, based on the parameter values of Table
\ref{tab:Baseline-Parameter}. Within the numerical simulation, 32
randomly generated chip sequences have been considered. Thereby, the
mean value (solid lines) and the one-sigma interval (dashed lines)
are depicted after smoothing.\label{fig:Phase_noise_lsd_analytic_vs_numeric}}
\end{figure}

Typical types of pulse modulation schemes are binary phase shift
keying (BPSK) and binary offset carrier (BOC). In this discussion,
we will restrict ourselves specifically to \mbox{BPSK-R} (where the
R indicates a rectangular pulse modulation) and sine-phased BOC(m,n),
with $m/n\in\mathbb{N}$. For a \mbox{BPSK-R} modulation with chip
period $T_{c}$, the PSD is given by: $|P_{\text{BPSK}}(f)|^{2}/T_{c}=T_{c}\,\text{sinc}^{2}(\pi fT_{c})$,
with $\text{sinc}(x)=\sin(x)/x$. Notably, this function exhibits
a maximum at the origin, i.e. at frequencies not being filtered at
the phase readout, cf. gray graph in Fig. \ref{fig:Phase_noise_lsd_analytic_vs_numeric}
a). Inserting the PSD of the pulse modulation into  \ref{eq:phase_noise_density_approx}
and  \ref{eq:variance_approx} results in:
\begin{align}
S_{\text{BPSK-R}}(f) & =\Pi_{-B_{\text{F}}^{\text{p}},B_{\text{F}}^{\text{p}}}(f)m_{\text{prn}}^{2}T_{c}\text{sinc}^{2}(\pi fT_{c}),\label{eq:S_BPSK}\\
 & \approx\Pi_{-B_{\text{F}}^{\text{p}},B_{\text{F}}^{\text{p}}}(f)m_{\text{prn}}^{2}T_{c},\label{eq:S_BPSK_approx}\\
\sigma_{\text{BPSK-R}}^{2} & =m_{\text{prn}}^{2}\int_{-B_{\text{F}}^{\text{p}}}^{B_{\text{F}}^{\text{p}}}T_{c}\text{sinc}^{2}(\pi fT_{c})\text{d}f,\label{eq:var_BPSK}\\
 & \approx2m_{\text{prn}}^{2}B_{\text{F}}^{\text{p}}T_{c}.\label{eq:var_BPSK_approx}
\end{align}
Thereby, the sine cardinal has been approximated using a Taylor expansion
according to $\text{sinc}(x)\approx1$, as $B_{\text{F}}^{\text{p}}\ll1/T_{c}$.
Importantly, the PSD of the phase noise exhibits a constant value
at low frequencies. Thus, the phase noise can only be reduced via
the chip period and the modulation index. However, when reducing
the modulation index to a level where the residual phase noise becomes
acceptable, the code tracking and associated ranging error, discussed
in the subsequent section, may become in-acceptably large. Similarly,
a smaller chip period may shift large parts of the spectral energy
of the modulation outside the receiver measurement bandwidth. This
situation can however be improved when applying modulation schemes
such as BOC, as shown hereafter.\\
First introduced by John Betz, BOC(m,n) is characterized by a square
sub-carrier modulation of the chips \cite{Betz2001}. The frequency
$f_{\text{sc}}$ of the sub-carrier is expressed by the index $m=f_{\text{sc}}/f_{\text{ref}}$,
where $f_{\text{ref}}$ represents a reference frequency. The second
index $n=f_{\text{c}}/f_{\text{ref}}$ defines the chip rate $f_{c}$.
As we restricted the signal to exhibit a constant pulse modulation
$p(t)$, cf. (\ref{eq:input_signal}), it is necessary to have a sub-carrier
multiple of the chip rate, yielding $m/n\in\mathbb{N}$. With no
loss of generality, we set $f_{c}=f_{\text{ref}}$, expressed as $n=1$.
Following these presumptions, the PSD of the sine-phased BOC(m,1)
is given by $|P_{\text{BOC(m,1)}}(f)|^{2}/T_{c}=T_{c}\,\text{sinc}^{2}(\pi fT_{c})\tan^{2}\left(\pi fT_{c}/(2m)\right)$
\cite{Betz2001}. In strong contrast to BPSK, the peak of the PSD
is shifted away from the origin, to $f\approx\pm f_{\text{sc}}$.
Moreover, the power contribution at the origin reads zero, cf. blue
graph in Fig. \ref{fig:Phase_noise_lsd_analytic_vs_numeric} a). Finally,
performing similar approximations as for the BPSK modulation results
in a phase noise PSD and variance of:
\begin{align}
S_{\text{BOC(m,1)}}(f) & =\Pi_{-B_{\text{F}}^{\text{p}},B_{\text{F}}^{\text{p}}}(f)m_{\text{prn}}^{2}T_{c}\text{sinc}^{2}(\pi fT_{c})\tan^{2}\left(\dfrac{\pi fT_{c}}{2m}\right),\label{eq:S_BOC}\\
 & \approx\Pi_{-B_{\text{F}}^{\text{p}},B_{\text{F}}^{\text{p}}}(f)m_{\text{prn}}^{2}T_{c}^{3}\left(\dfrac{\pi f}{2m}\right)^{2},\label{eq:S_BOC_approx}\\
\sigma_{\text{BOC(m,1)}}^{2} & =m_{\text{prn}}^{2}\int_{-B_{\text{F}}^{\text{p}}}^{B_{\text{F}}^{\text{p}}}T_{c}\text{sinc}^{2}(\pi fT_{c})\tan^{2}\left(\dfrac{\pi fT_{c}}{2m}\right)\text{d}f,\label{eq:var_BOC}\\
 & \approx\left(\dfrac{\pi m_{\text{prn}}}{m}\right)^{2}\dfrac{\left(B_{\text{F}}^{\text{p}}T_{c}\right)^{3}}{6}.\label{eq:var_BOC_approx}
\end{align}
Strikingly, and in strong contrast to BPSK modulation, the phase noise
PSD for BOC(m,1) reads zero at the origin and increases quadratically.
Consequently, $S_{\text{BOC(m,1)}}(f)$ exhibits a maximum at $f=B_{\text{F}}^{\text{p}}$.
The ratio at this maximum between BOC(m,1) and BPSK modulation is
$\left(B_{\text{F}}^{\text{p}}T_{c}\pi/2m\right)^{2}$. Thus as long
as $B_{\text{F}}^{\text{p}}T_{c}\ll1$, BOC(m,1) modulation exhibits
superior noise performance compared to BPSK modulation. A similar
conclusion holds for the variance. \\
The effect of the modulation becomes especially noticeable when considering
parameter values base-lined for LISA, see Table \ref{tab:Baseline-Parameter}.
In the context of LISA, two modulation schemes have been extensively
discussed in various publications: BPSK and Manchester encoding \cite{Wand2007,sutton2010laser,Esteban:11,Delgado2012, Sutton_2013}.
Within the ongoing spectral analysis, Manchester encoding can equivalently
be represented as a BOC(1,1). Moreover, in this context, the phase
read-out accuracy is analyzed in terms of displacement noise. For
this purpose, the phase noise is considered as linear spectral density
(LSD), obtained via the square root of the PSD. Converting the phase
noise into displacement noise by multiplication with the conversion
factor $\lambda/(2\pi)$, where $\lambda$ denotes the wavelength
at the heterodyne detection, yields a single-sided LSD $L(f)=\sqrt{2S(f)}\,\lambda/(2\pi)$
in $\text{m}/\sqrt{\text{Hz}}$, which for BPSK modulation exceeds
the BOC(1,1) LSD by five orders of magnitude.\\
This significant difference is verified by a numerical PLL simulation,
set up according to the generic model depicted in Fig. \ref{fig:Generic-Receiver-Design}
and parameter values listed in Table \ref{tab:Baseline-Parameter}.
Thereby, 32 randomly generated PRN sequences, modulated either via
BPSK (gray lines) or BOC(1,1) (blue lines) have been considered. Fig.
\ref{fig:Phase_noise_lsd_analytic_vs_numeric} b) illustrates the
single-sided LSD $L(f)$ of the displacement noise, where the mean
value (solid lines) and the one-sigma interval (dashed lines) are
depicted after smoothing. Both numerical simulations agree well with
the respective analytical model, exhibiting a constant slope for the
LSD of the BPSK modulation and a linear one for the BOC(1,1) modulation.
Deviations from the analytical model are attributed to the finite
chip sequence length and are found to vanish for infinitely long sequences.
Importantly, these results manifest the superior phase noise performance
of the BOC(1,1) modulation. On the other hand, they exclude BPSK modulation
for the given set of parameter values for applications requiring pico-meter
noise levels at the phase readout. 

\begin{figure}[!t]
\centering{}\includegraphics[width=1\linewidth]{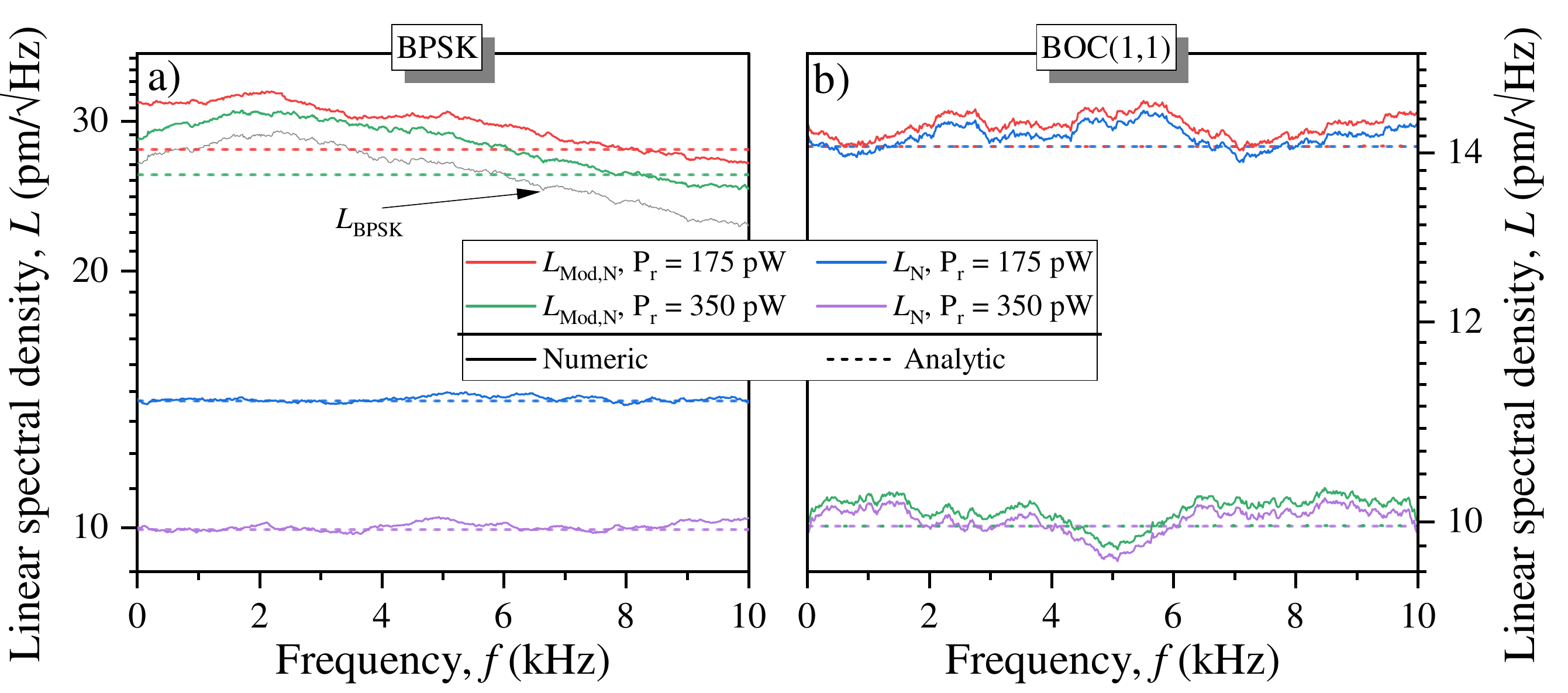}\caption{Panels a) and b) illustrate the single-sided displacement noise LSD
of a numerical PLL simulation for a LISA representative signal considering
BPSK and BOC(1,1) modulation, respectively. The incoming signal of
the red and green graph consists of the science signal, the modulation
(Mod) and the relevant LISA noise contributions (N), while modulation
is absent in the incoming signal of the blue and purple graph. Both
pairs differ in the power level $P_{\text{r}}$ of the incoming signal.
In addition to the simulated values represented by the solid lines,
analytical models are shown in the same color by the dashed curves.
Finally, the single-sided displacement noise LSD resulting only from
BPSK modulation (i.e. without external noise) is shown by the gray
graph as a reference in panel a). The simulation is based on the parameter
values of Table \ref{tab:Baseline-Parameter}. Within the numerical
simulation, 16 randomly generated chip sequences have been considered.
Thereby, the mean value is depicted after smoothing.\label{fig:Phase_noise_lsd_incl_noise}}
\end{figure}

\textcolor{blue}{}Until now, the analysis has focused on an ideal
heterodyne signal as described in (\ref{eq:input_signal}). This approach
facilitated a performance evaluation based exclusively on architectural
design decisions. However, in practical scenarios, the incoming signal
often deviates significantly from this ideal model. Consequently,
in the following section, we contextualize our analysis with a focus
on representative signals, particularly those encountered in the LISA
mission.

The heterodyne signal of LISA is given by the beatnote of the received
weak signal from a remote spacecraft and a local laser beam, having
powers of $P_{\text{r}}=$ 350 pW and $P_{\text{lo}}\sim$ 1 mW, respectively.
These signals interfere at a photodiode, where based on the responsivity
(cf. Table \ref{tab:Baseline-Parameter}), the incident power is converted
into a photocurrent. While redundancy and averaging concepts for
photodiode segments have been proposed \cite{Morrison:94,Morrison:94_2,Heinzel2020},
these aspects are beyond the scope of this paper, and the following
analysis considers a single photodiode.

Following the detection principle of LISA, the nominal signal is affected
by several noise contributions, see \cite{LISAPRFMODEL}. Tilt-to-length
(TTL) coupling noise, arising from the coupling between pupil alignment
offsets and spacecraft jitter, is removed in post-processing. Further
neglecting stray light coupling noise (very small) and phasemeter
internal noise, which depends on the specific hardware and can be
made sufficiently small ($\sim$1 pm/sqrt(Hz)), leaves two dominating
contributors, namely shot noise and laser intensity noise. These
noise contributions have been incorporated into the incoming signal
model applied for the numerical PLL simulations, thus providing a
LISA representative signal. Relevant parameter values have been taken
from \cite{Wissel2023}. The resulting displacement noise spectra
averaged over 16 randomly generated PRN sequences are depicted in
Fig. \ref{fig:Phase_noise_lsd_incl_noise} a) and b) after smoothing.
In absence of the PRN modulation, the displacement noise LSD $L_{\text{N}}$
exhibits a flat spectrum, cf. purple curves Fig. \ref{fig:Phase_noise_lsd_incl_noise}
a) and b). These spectra are accurately described as uncorrelated
summation of analytical formulations for shot noise \cite{Delgado2012}
and laser intensity noise \cite{Hechenblaikner:13,Wissel2022,Wissel2023}.
The same characteristics apply when the power $P_{\text{r}}$ of the
incoming laser is halved, as shown by the blue curves in Fig. \ref{fig:Phase_noise_lsd_incl_noise}
a) and b). Comparing these results to those obtained for an incoming
signal also comprising a PRN-modulated signal component, indicated
as $L_{\text{Mod,N}}$, substantiates the relevance of the choice
of modulation in the context of LISA.

In Fig. \ref{fig:Phase_noise_lsd_incl_noise} a), we illustrate the
impact of BPSK modulation. The phase noise is strongly dominated by
the modulation, irrespective of the power levels considered, cf. green
and red graphs. The influence becomes clearly evident by comparison
with the gray graph $L_{\text{BPSK}}$, representing phase noise that
derives solely from the BPSK modulation. In sharp contrast, Fig.
\ref{fig:Phase_noise_lsd_incl_noise} b) demonstrates the effect of
BOC(1,1) modulation, which contributes minimally to the overall phase
noise. These findings are consistent with the analytical models,
c.f. dashed lines, while deviations are again attributed to the finite
PRN chip sequence length. Importantly, these results verify the substantial
influence of the modulation scheme on the phase noise performance
for a LISA-representative environment.\textcolor{blue}{}\\
\textcolor{blue}{}Finally, without data transmission, the signal
consists of only periodical spreading sequences, which yield in the
frequency domain a comb around the origin spaced by the inverse of
the code sequence periodicity. Since $B_{\text{F}}^{\text{p}}\ll1/T_{s}$,
only the peak at the origin may affect the result, which in the case
of a BOC(m,1) modulation is suppressed, due to the symmetry properties
of the pulse.

\section{Code tracking and ranging performance\label{sec:Code-tracking-and}}

\subsection{Principle of DLL\label{subsec:Principle-of}}

After carrier wiping, a DLL relying on the principle of a non-coherent
early-late discriminator, and extensively used in GNSS applications,
is capable of tracking the code according to (\ref{eq:input_signal}).
Thereby, the principle of the delay detector, see Fig. \ref{fig:Generic-Receiver-Design},
relies on two local code replicas -- forwarded and delayed in time,
where the particular delay of the code replicas, i.e. the early-late
spacing, depends strongly on the modulation technique \cite{Betz2000DesignAP, Van1992TheoryAP}.
These replicas are correlated with the incoming signal over one symbol
period, followed by a squared magnitude operation, to suppress data
polarity. Finally, the difference between early and late correlation
yields a value on the so-called S-curve $S(\varepsilon)$. Thereby,
the argument $\varepsilon$ indicates the time shift between the replica
and the incoming signal. The gray curve in Fig. \ref{fig:S-curve}
b) illustrates a S-curve for an exemplary BPSK-modulated incoming
chip sequence which equals its replica. A time segment of the incoming
signal, and thus of the replica, is depicted in Fig. \ref{fig:S-curve}
a) by the gray curve. For this ideal incoming signal, the zero crossing
corresponds to perfect alignment between replica and incoming signal,
and is usually considered as the tracking point of the loop, with
a tracking range corresponding to the linear regime around the zero
crossing. In addition, the blue graph in Fig. \ref{fig:S-curve} b)
depicts the S-curve for a BOC(1,1)-modulated signal and its identical
replica, whose time segment is portrayed in Fig. \ref{fig:S-curve}
b), also by the blue graph. In contrast to the BPSK-modulated S-curve,
there are additional stable tracking points, represented by additional
zero-crossings within a region of positive slope. These play a key
role during loop (re-)acquisition but are not further elaborated in
the following discussion. Moreover, the linear range is reduced,
due to the necessarily smaller early-late spacing, cf. Table \ref{tab:Baseline-Parameter}.
Irrespective of the modulation, within the linear range, the time
shift between the replica and the incoming signal is obtained via
the division of the S-curve value by the constant slope $m=\partial_{\varepsilon}S$
in this regime. In this context, the slope is usually represented
as a discriminator gain $d_{\text{g}}=1/m$ \cite{Betz2000DesignAP}.
Finally, this shift serves as an input to the low-pass filter, which
estimates the chip rate used as input to the PRN code generator.

\subsection{Signal modeling\label{subsec:Signal-modeling}}

\begin{figure}[!t]
\begin{centering}
\includegraphics[width=1\linewidth]{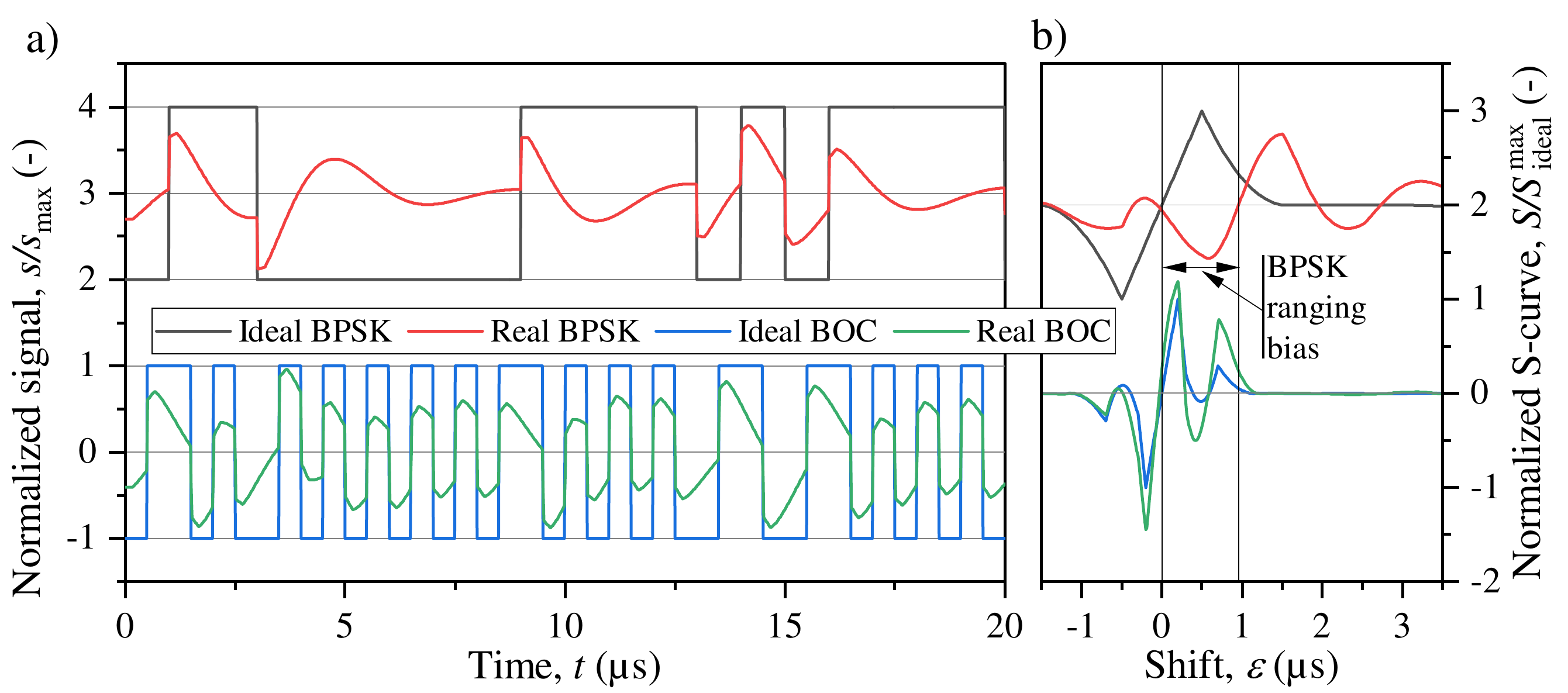}
\par\end{centering}
\centering{}\caption{Panel a) illustrates exemplary time segments of BPSK- and BOC(1,1)-modulated
chip sequences. Thereby, the signals $s(t)$ are normalized to their
maximum value $s_{\text{max}}$ and represent an ideal, i.e. unfiltered,
incoming wide-band signal (gray graph) and the real, i.e. filtered
by the PLL, incoming signal (red graph) for BPSK modulation. Similarly,
for BOC(1,1) modulation unfiltered and filtered signals are depicted
by the blue and green graphs, respectively. In addition, the BPSK-modulated
signals are shifted for clarity by $s/s_{\text{max}}=3$. The corresponding
S-curves for the replica and the ideal signal (gray curve for BPSK
and blue curve for BOC), as well as for the replica and the real signal
(red curve for BPSK and green curve for BOC) are presented in panel
b). Thereby, the S-curves are normalized to the maximum value of the
ideal S-curve. The S-curves based on the real signals differ in shape
and also in the zero crossing of the linear range compared to the
S-curves based on the ideal signals. The shift of the zero-crossing
is thereby denoted as ranging bias, exemplary indicated for BPSK modulation
by the black double arrow. Again, the BPSK-modulated S-curves are
offset for clarity by $S/S_{\text{ideal}}^{\text{max}}=2$.\label{fig:S-curve}}
\end{figure}

In strong contrast to the ideal, i.e. unfiltered, case stated in section
\ref{subsec:Principle-of}, the architecture depicted in Fig. \ref{fig:Generic-Receiver-Design},
not only wipes the carrier but also affects the code sequences. Based
on standard control theory the code sequences are filtered by the
impulse response of the error transfer function $e(t)$ of the PLL,
yielding the signal $\tilde{s}(t)=e(t)\ast m_{\text{prn}}\sum_{j=-\infty}^{\infty}d_{j}\sum_{i=0}^{N-1}c_{i}p(t-iT_{c}-jNT_{c})$,
at the input of the DLL. Taking into account the high-pass filter
behavior of the error transfer function and the PSD of the two modulation
methods, cf. Fig. \ref{fig:Phase_noise_lsd_analytic_vs_numeric} a),
we find that BPSK-modulated sequences are significantly more distorted
than BOC(1,1)-modulated sequences. This behavior is illustrated by
the red and green curves of Fig \ref{fig:S-curve} a), for BPSK and
BOC(1,1) modulation, respectively, which constitute the filtered signals
used as input to the DLL. Nonetheless, both signals are characterized
by overshoots at the beginning of a chip value transition and strong
damping toward the end of the chip. As a consequence, the resulting
S-curves, based on the correlation of the filtered signal with an
unfiltered replica, differ significantly from the ideal case. In particular,
for BPSK modulation, cf. red graph in Fig \ref{fig:S-curve} b), the
distortion leads to additional stable tracking points. Besides the
shape also the zero crossing of the linear range, i.e. the primary
tracking point, is shifted by nearly one chip period, which will be
referred to as ranging bias \cite{Betz2002DesignAP}. For BOC(1,1)
modulation, alternation in shape and zero crossing are moderate, cf.
green graph in Fig \ref{fig:S-curve} b).

\begin{figure}[!t]
\begin{centering}
\includegraphics[width=1\linewidth]{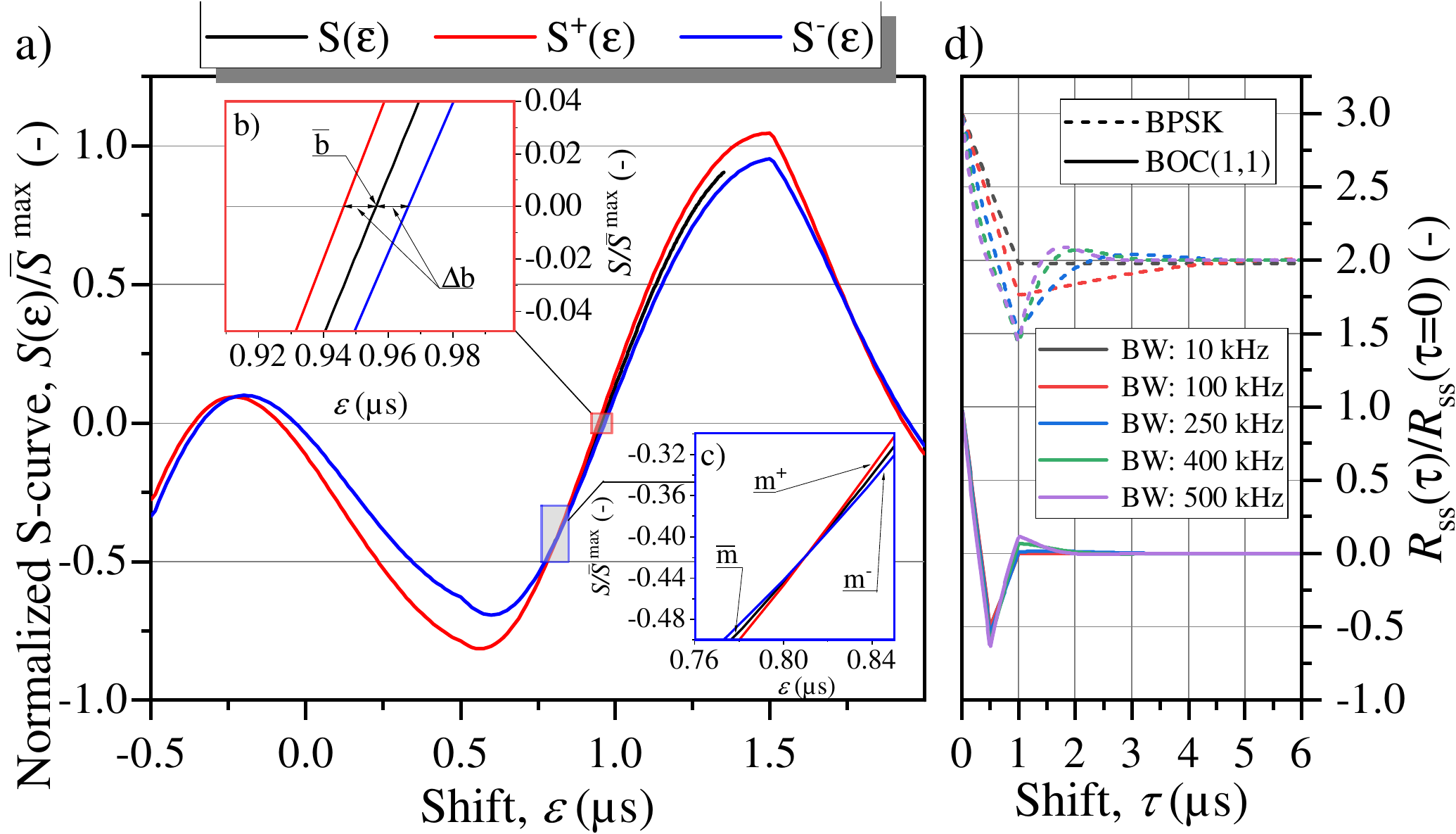}
\par\end{centering}
\centering{}\caption{In panel a), the S-curves $S^{+}$(red curve) and $S^{-}$ (blue curve),
denoting whether the current and previous data symbol exhibit the
same ($+$) or opposite ($-$) value, are depicted for an exemplary
BPSK-modulated chip sequence. Moreover, in the linear range the mean
S-curve, corresponding to the mean shift $\bar{\varepsilon}$, is
illustrated by the black curve. All S-curves are normalized to the
maximum of the mean value $\bar{S}(\varepsilon)=(S^{+}(\varepsilon)+S^{-}(\varepsilon))/2$.
Panel b) shows the area at the zero crossing, indicating the parameters
$\bar{b}$ and $\Delta b$. Finally, panel c) portrays the slope variation
between the S-curves $S^{+}$ and $S^{-}$, indicating the parameters
$\bar{m}$, $m^{+}$ and $m^{-}$. Panel d) shows the normalized autocorrelation
$R_{ss}$ for an uncorrelated chip sequence filtered by a high-pass,
representing the error transfer function of the PLL, at different
filter bandwidths (BW). The results are depicted for BPSK- and BOC(1,1)-modulated
sequences, indicated by the dashed and solid lines, respectively.
Thereby, the former is offset by $R_{ss}(\tau)/R_{ss}(0)=2$ for clarity.
\label{fig:two-S-curve-for-chip-flip-and-no-chip-flip}}
\end{figure}

Moreover, due to the convolution operation of the filtering, i.e.
due to the filter memory, the symbols not only differ in sign, which
is well accounted for by the magnitude (squared) operation but rather
they depend on the input of the previous symbols. Consequently,
the corresponding S-curve varies over time, yielding a ranging bias
variation.\\
As a matter of fact, analyzing an uncorrelated chip sequence after
exposure to a high-pass filter, representative for the error transfer
function of the PLL, reveals a correlation time limited to several
chip periods. This behavior is observed for BPSK and BOC(1,1) modulation
for a bandwidth (BW) of 10 - 500 kHz, cf. Fig. \ref{fig:two-S-curve-for-chip-flip-and-no-chip-flip}
d), which appears as relevant PLL bandwidth range considering a chip
rate of 1 MHz. These findings exclude the persistence of correlation
over more than one symbol length (64 chips, cf. Table \ref{tab:Baseline-Parameter}).
Therefore, a symbol and its modulated data bit can have at most an
impact on the processing of the succeeding symbol and data bit (memory
effect). 
In addition, these findings are affirmed by numerical analysis based
on randomly generated PRN sequences and parameter values stated in
Table \ref{tab:Baseline-Parameter}, revealing that variations of
the S-curve are restricted to $S^{+}$ and $S^{-}$, depending on
whether the current and previous data symbols exhibit the same ($+$)
or opposite ($-$) value. These S-curves, depicted in Fig. \ref{fig:two-S-curve-for-chip-flip-and-no-chip-flip}
a) for an exemplary BPSK-modulated PRN sequence, promote the introduction
of a mean S-curve $S(\bar{\varepsilon})$ in the linear range, exhibiting
a mean ranging bias $\bar{b}$ as portrayed in Fig. \ref{fig:two-S-curve-for-chip-flip-and-no-chip-flip}
b). Importantly, $\bar{\varepsilon}=[\varepsilon(S^{+})+\varepsilon(S^{-})]/2$
indicates the mean value based on the x-axis, which can be found via
interpolation of $S^{\pm}(\varepsilon)$ in the linear range of $S^{+}$
and $S^{-}$. This leads to a common offset $\Delta b$ between the
mean S-curve $S(\bar{\varepsilon})$ and $S^{\pm}$ at the zero crossing.
Consequently, the ranging bias can be modeled as:
\[
b(t)=\Delta b\sum_{n=-\infty}^{\infty}\hat{f}_{n}\Pi_{nT_{s},(n+1)T_{s}}(t)+\bar{b}.
\]
Thereby, the variable $\hat{f}_{n}\in\left\{ \pm1\right\} $ expresses
the similarity of the current and the previous data symbol, according
to:
\[
\hat{f}_{n}=\begin{cases}
1 & d_{n}=d_{n-1}\\
-1 & \text{else}.
\end{cases}
\]
The boxcar function $\Pi_{nT_{s},(n+1)T_{s}}(t)$ indicates the variation
in the symbol period $T_{s}$. It shall be emphasized, that the mean
ranging bias $\bar{b}$ and the deviation $\Delta b$, strongly depend
on the specific code sequence and thus need to be determined numerically.\\
As long as the mean ranging bias is in the linear range of the S-curve
it only constitutes the tracking point of the DLL, which can be accounted
for by means of calibration and is thus omitted for further discussion.
In contrast, the variation of the ranging bias confines the accuracy
of the tracking loop. In order to identify the corresponding noise
contribution, one can use (\ref{eq:noise_term_stochastic}) which
describes the stochastic noise of BPSK modulation for the PLL, and
apply it instead to the DLL analyzed in this section, by establishing
a correspondence between the following parameters: $m_{\text{prn}}\rightarrow\Delta b$,
$\hat{X}_{n}\rightarrow\hat{f}_{n}$, $T_{c}\rightarrow T_{s}$, leading
to a noise spectral density of
\begin{align*}
N_{b}(f) & =\Delta b^{2}T_{s}\text{sinc}^{2}(\pi fT_{s}).
\end{align*}
This noise spectral density is low-pass filtered within the DLL, followed
by a read-out filter. Approximating both filters as an ideal low-pass
filter, the variance at the readout is given by:
\begin{align}
\sigma_{b}^{2} & =\Delta b^{2}\int_{-B_{\text{F}}^{\text{d}}}^{B_{\text{F}}^{\text{d}}}T_{s}\text{sinc}^{2}(\pi fT_{s})\text{d}f,\nonumber \\
 & \approx2\Delta b^{2}B_{\text{F}}^{\text{d}}T_{s}.\label{eq:variance_ranging_error}
\end{align}
Thereby, $B_{\text{F}}^{\text{d}}$ indicates the read-out filter
bandwidth of the DLL, which is assumed to be much smaller than the
inverse of the symbol period and smaller or equal to the bandwidth
of the DLL low-pass filter. The error of the DLL is usually considered
in terms of a ranging error. Thus, the code-tracking error $\sigma_{\text{r,a}}$
of this semi-analytical model will be defined as $\sigma_{\text{r,a}}=c\sigma_{b}$,
where $c$ denotes the speed of light. Similar expressions for the
code-tracking error as stated in (\ref{eq:variance_ranging_error})
have been found for alternative DLL implementations \cite{BetzGen2009}.

\subsection{Application and numerical verification}

\begin{figure}[!t]
\begin{centering}
\includegraphics[width=1\linewidth]{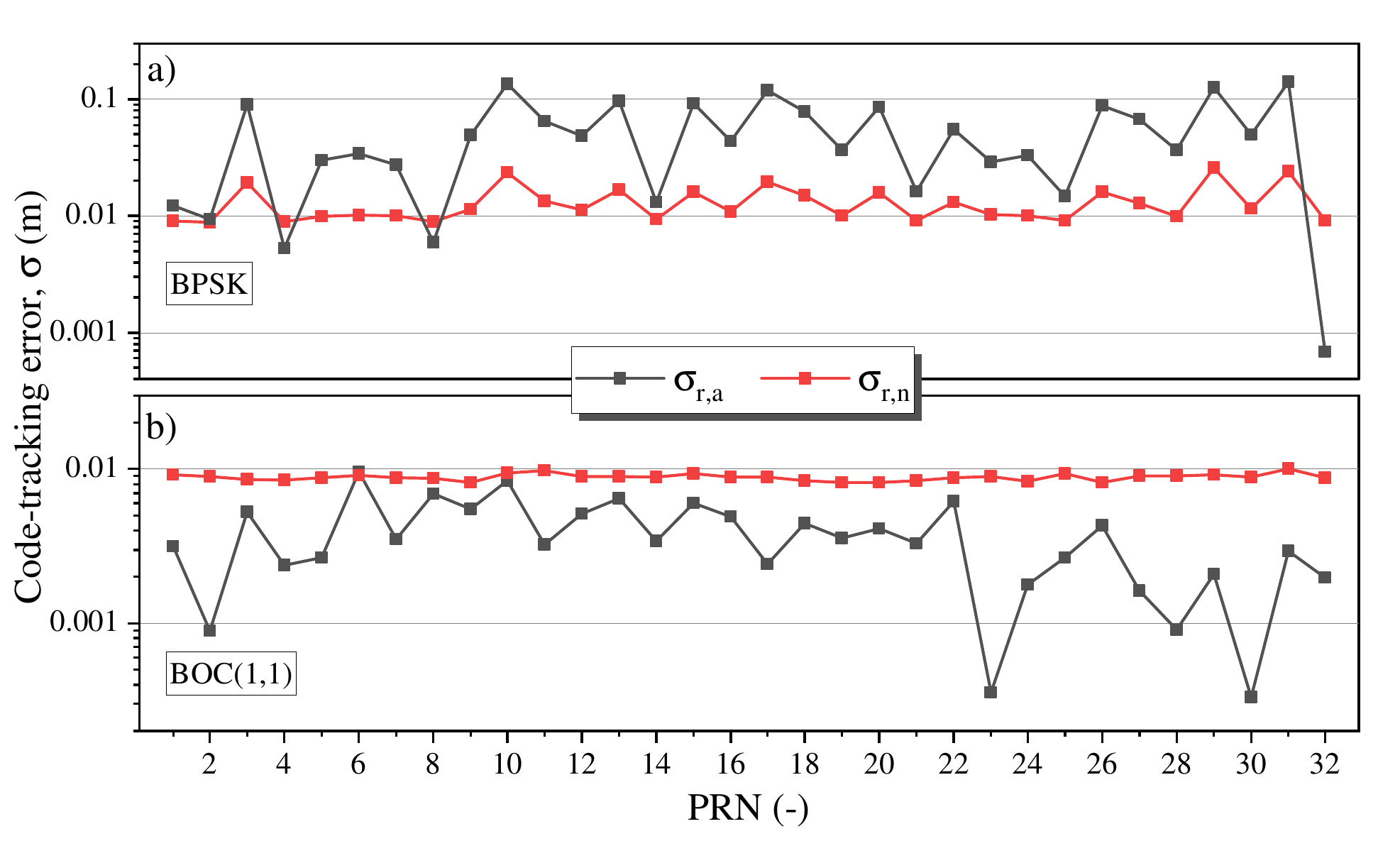}
\par\end{centering}
\centering{}\caption{The code-tracking error of a numerical DLL simulation (red data points)
and a semi-analytical model (gray data points) is depicted for 32
randomly generated PRN sequences. In panel a) results are displayed
for BPSK-modulated PRN sequences, while panel b) illustrates the outcome
for BOC-modulated sequences. \label{fig:two-S-curve-vs-DLL-BPSK-vs-BOC}}
\end{figure}

Analogous to carrier tracking, also for code tracking numerical simulations
have been conducted comparing BPSK and BOC(1,1) modulation schemes
and verifying the semi-analytical model for the code-tracking error
$\sigma_{\text{r,a}}$, cf. (\ref{eq:variance_ranging_error}). Besides
the baseline parameters as specified in Table \ref{tab:Baseline-Parameter},
a set of 32 randomly generated PRN sequences has been considered.
In terms of the semi-analytical code-tracking error $\sigma_{\text{r,a}}$,
BOC(1,1) modulation displays a superior performance, differing by
around one order of magnitude compared to the BPSK modulation, cf.
gray data points in Fig. \ref{fig:two-S-curve-vs-DLL-BPSK-vs-BOC}
a) and b). This result is expected: as explained in section \ref{sec:Carrier-tracking-and},
the BPSK modulation holds its peak spectral energy at the origin,
leading to maximum damping due to the high-pass filter behavior of
the error transfer function of the PLL. On the other hand, the spectral
energy of the BOC(1,1) modulation is shifted away from the carrier,
yielding less spectral confinement, cf. Fig. \ref{fig:Phase_noise_lsd_analytic_vs_numeric}
a), and hence less distorted PRN sequences, see Fig. \ref{fig:S-curve}
a) \cite{Delgado2012}. Consequently, the S-curve of the BOC(1,1)
modulation exhibits a smaller ranging bias variation $\Delta b$ as
well as an absolute ranging bias $\bar{b}$ that is smaller by approximately
two orders of magnitude.\\
The code-tracking error $\sigma_{\text{r,n}}=c\sqrt{\text{var}\left(\varepsilon\right)}$
of the numerical DLL simulation is based on the variance $\text{var}\left(\right)$
of the time shift $\varepsilon$, once the DLL has been settled.
Interestingly, a significant deviation between modeled and simulated
code-tracking errors is visible irrespective of the modulation scheme.
For BOC(1,1) modulation the numerical code-tracking error is nearly
constant over the set of PRN sequences, exhibiting a value of $\sigma_{\text{r,n}}\approx$
9 mm, cf. red data points in Fig. \ref{fig:two-S-curve-vs-DLL-BPSK-vs-BOC}
b). Moreover, a correlation to the semi-analytical model is not evident
at first glance. Although the code-tracking error of the model and
the simulation are clearly correlated for BPSK modulation, see red
data points in Fig. \ref{fig:two-S-curve-vs-DLL-BPSK-vs-BOC} a),
deviations are still apparent in terms of the amplitude of the code-tracking
error. These deviations are attributed to two effects: (i) the granularity
of the PRN code generator, and (ii) the varying slopes of the S-curves.
\\
Regarding effect (i), the resolution of the PRN code generator is
confined by the sampling rate of the incoming chip sequence. For BOC(1,1)
modulation, the sampling rate of 80 MHz cannot resolve the ranging
bias variations \cite{EuringerOptMetr2023}. Consequently, simulations
exhibit equal code-tracking errors, irrespective of the PRN sequence.
The increased ranging bias variation for the BPSK modulation leads
to a noticeable influence of the bias variation and hence establishes
a correlation between the semi-analytical model and the simulation.
Still, the variations are not fully resolved by the sampling rate.\\
\textcolor{brown}{}In addition, effect (ii) is caused by the slope
$m$ of the S-curves, necessary for the detection of the time shift
$\varepsilon$. The DLL deduces the time shift based on the discriminator
gain and thus implies a constant S-curve slope, as delineated in section
\ref{subsec:Principle-of}. Because the slopes $m^{\pm}$ in the linear
range of the S-curves $S^{\pm}$ are not identical, see Fig. \ref{fig:two-S-curve-for-chip-flip-and-no-chip-flip}
c), the TOA estimation induces an error, yielding an additional deviation
from the proposed semi-analytical model.\\
Further insight into effect (i) is gained through the analytical model
of \cite{EuringerOptMetr2023} which investigates the ranging error
resulting from the granularity of the PRN code generator. This model
reveals that for the considered parameter values, the dominant error
results from the granularity of the PRN code generator, i.e. the sampling
error.\textcolor{brown}{}\\
In addition, the ranging performance has been evaluated for a LISA
representative signal that is degraded by the external shot noise
and laser intensity noise (see also subsection \ref{subsec:Application-and-numerical}).
These noise contributions lead to ranging errors that exceed the error
caused by the granularity of the PRN code generator, see above effect
(i). In this case, the ranging error for BOC(1,1) (around several
centimeters, depending on the specific code sequence) is around four
times smaller than for BPSK, which reveals the superiority of BOC(1,1)
encoding for LISA. Comparable findings have been reported in \cite{sutton2010laser,Delgado2012}.\textcolor{blue}{}
Nonetheless, both modulation schemes are capable of achieving sub-meter
ranging errors, which is considered sufficient \cite{EESA}.

At this point, it shall be emphasized, that due to the specific TOA
detection, pure analytical analysis for the DLL is much more complex
compared to the PLL. In particular, this applies to common simplifications,
e.g. performed by Betz \cite{BetzGen2009} (or further ones not shown
here), which are not applicable in this context. These findings reinforce
the necessity of numerical analysis for distinct code sequences and
receiver architectures.

\section{Conclusion}

This paper revealed the compelling influence of the modulation scheme
on the performance of sequential carrier- and code-tracking receiver
architectures foreseen for future space-borne metrology systems.

A generic sequential PLL--DLL design including a representative signal
consisting of a carrier modulated by code sequences has been introduced,
enabling a novel analysis of the performance losses resulting exclusively
from the architecture itself. Thereby, carrier- and code-tracking
analyses have been conducted separately. In the former case a generic
model has been introduced, exploiting the distinct parameter range
and estimating the phase noise for an arbitrary but periodic modulation
scheme. Thereby, the PSD of the pulse modulation within the read-out
bandwidth has been identified as the main driver for phase noise.
This model, subjected to BPSK and BOC(1,1) modulation revealed the
superior phase noise performance of the latter. Moreover, it excluded
BPSK as a modulation scheme for space-borne metrology systems demanding
pico-meter noise levels at the phase readout, considering the stated
set of parameter values. Finally, these results have been verified
by numerical PLL simulations, which agreed well with the analytical
model regardless of the modulation scheme.

Analysis of the code tracking has been focused on the TOA estimation,
taking into account the concept of the S-curve. Thereby, a varying
S-curve due to the PLL filtering of the incoming signal has been observed,
differing in shape and zero crossing from the ideal case. Remarkably,
analyses revealed that variations of the S-curve can be reduced to
two cases, depending on whether the current and previous data symbols
exhibit the same or opposite value, enabling a similar mathematical
approach for the code tracking as used before for the phase noise.
Finally, the model was compared with numerical DLL simulations. Differences
became apparent, which were primarily attributed to the granularity
of the PRN code generator. While both modulation schemes exhibited
sub-meter ranging errors, BOC(1,1) modulation surpasses but at least
equals the performance of the BPSK modulation.\\
Sequential carrier and code tracking architectures are thus in principle
capable of serving as receivers for high-precision space-borne measurement
systems, but performance is significantly affected by the modulation
scheme. While the analysis was restricted to one data symbol per PRN
sequence, it can easily be extended to analyses of PRN sequences exhibiting
multiple data symbols, facilitating higher data rates. 


%



\section*{Acknowledgment}
The authors thank T. Ziegler and S. Delchambre for their support and fruitful discussions.

\ifCLASSOPTIONcaptionsoff
  \newpage
\fi



%
\bibliographystyle{IEEEtran_eu}
\bibliography{references_tim}

%








\end{document}